\documentclass[useAMS,amssymb,epsfig,usegraphicx,twocolumn,usenatbib]{mn2e}
\def\aj{AJ}
\def\araa{ARA\&A}
\def\nat{Nature}
\def\apj{ApJ}
\def\apjl{ApJ}
\def\aap{A\&A}
\def\mnras{MNRAS}
\def\nat{Nature}
\def\na{NewA}

\usepackage[usenames,dvips]{color}

\newif\ifAMStwofonts

\makeatother

\title[Secular evolution and cylindrical rotation]
{Secular evolution and cylindrical rotation in boxy/peanut bulges: impact of initially rotating classical bulges}

\author[Saha et al.]
{Kanak Saha\thanks{E-mail:saha@mpe.mpg.de}, \& Ortwin Gerhard\\
Max-Planck-Institut f\"ur Extraterrestrische Physik, Giessenbachstraße, D-85748 Garching, Germany}
 
\begin{document}

\date{Accepted xxxx Month xx. Received xxxx Month xx} 
\pagerange{\pageref{firstpage}--\pageref{lastpage}} \pubyear{2009}
\maketitle

\label{firstpage}

\begin{abstract}
Boxy/peanut bulges are believed to originate from galactic discs through 
secular processes. A little explored question is how this 
evolution would be modified if the initial disc was assembled around a 
preexisting classical bulge. Previously we showed that a low-mass initial 
classical bulge (ICB), as might have been present in Milky Way-like galaxies,
 can spin up significantly by gaining angular momentum from a bar formed through 
disc instability. Here we investigate how the disc instability and the kinematics
of the final boxy/peanut (BP) bulge depend on the angular momentum of such a low-mass 
ICB. We show that a strong bar forms and transfers angular momentum to the ICB in all 
our models. However, rotation in the ICB limits the emission of the bar's angular momentum, 
which in turn changes the size and growth of the bar, and of the BP bulge formed from 
the disc.

The final BP bulge in these models is a superposition of the BP bulge
formed via the buckling instability and the spun-up ICB. We find that
the long-term kinematics of the composite BP bulges in our simulations
is independent of the rotation of the ICB, and is always described by
cylindrical rotation.  However, as a result of the co-evolution between
bulge and bar, deviations from cylindrical rotation are seen during
the early phases of secular evolution, and may correspond to similar
deviations observed in some bulges. We provide a simple 
criterion to quantify deviations from pure cylindrical rotation, apply it 
to all our model bulges, and also illustrate its use for two galaxies: 
NGC7332 and NGC4570.

\end{abstract}


\section{Introduction}
\label{sec:introduc}
The formation of realistic disc galaxies and their evolution has
remained a challenging problem to the cold dark matter (CDM) paradigm of
galaxy formation \citep{WhiteRees1978}. Nevertheless, significant development
has occurred in this context in recent years.
Cosmological hydrodynamical simulations which include feedback and/or 
smooth accretion of cold gas through cosmic filaments show that an
exponential disc could have assembled either around a classical bulge formed
through mergers or as a bulgeless disc galaxy, and
survived through the hierarchical assembly \citep{Abadietal2003, Governatoetak2007,
Agertzetal2011, Brooketal2011, Brooketal2012}. 

Soon after their formation, the clumpy discs observed at
high-redshift \citep{Genzeletal2006} are believed to go through rapid
dynamical evolution where minor mergers, streams of
inflowing gas triggering burst of star formation and turbulence played dominant
roles. As these violent processes become less important and frequent
\citep{Hopkinsetal2010b,Lotzetal2011}, secular processes which operate on
longer time scales (typically $\sim 10$s of dynamical time scales) are thought to
play a significant role in the subsequent evolution of the galaxies
\citep{KormendyKennicut2004}. In the secular phase, the most
efficient way a disc galaxy evolves is through forming a bar via disc
instability which facilitates the redistribution of energy and angular
momentum between the disc, dark matter halo and 
classical bulge \citep{DebattistaSellwood2000,Athanassoula2003, Sahaetal2012}.
As the bar grows stronger, it goes through buckling instability and form
boxy/peanut bulges (hereafter, BP) as demonstrated in numerous N-body simulation
studies \citep{CombesSanders1981,PfennigerNorman1990, Rahaetal1991,
MV2004,Debattistaetal2006, Sahaetal2012}. BP bulges are seen in
nearly $50 \%$ of edge-on disc galaxies \citep{Luttickeetal2000} including our
Milky Way and the line-of-sight (LOS) stellar and gas kinematics in many
such barred edge-on galaxies show cylindrical rotation \citep{KI1982, 
Bureau1999,Falconbarrosoetal2006}, as has also been noted in many N-body 
simulations of barred galaxies \citep{Combesetal1990,SellwoodWilkinson1993,
Athanamisi2002,Sahaetal2012}. Stellar kinematics from the recently completed
Bulge Radial Velocity Assay (BRAVA) survey confirms the cylindrical rotation in the Galactic Bulge \citep{Howardetal2009, Kunderetal2012}.
Recent observational analyses, however, show deviations from cylindrical 
rotation in BP bulges, e.g., NGC5746, NGC1381
\citep{Williamsetal2011}. Deviations from cylindrical rotation is also
reported in BP bulges of simulated galaxies due to variation in the viewing angles
of the bar \citep{Combesetal1990,Athanamisi2002}. However, it is not 
clear whether the projection effect is the correct interpretation, or whether 
a physical mechanism is required to explain the deviations from cylindrical 
rotation observed in BP bulges.

BP bulges are not only formed out of pure axisymmetric discs but they can also form
in axisymmetric discs assembled around classical bulges \citep{Sahaetal2012}. 
Such classical bulges are presumably have
formed as a result of violent processes e.g., dissipative collapse 
\citep{Eggenetal1962} or mergers \citep{Baughetal1996}. In observation, one does
find both barred and unbarred galaxies with classical bulges \citep{Laurikainenetal2007}.
Classical bulges are indeed abundant in disc galaxies and they are known to rotate
\citep{KI1982,Cappellarietal2007}. In a previous paper, \cite{Sahaetal2012}
showed how an initially non-rotating low-mass classical bulge spun-up
during the secular evolution and discussed the possible implication it might
have on the final BP bulge. Such a low-mass classical bulge 
($\sim 8\%$ of disc mass) might be present in the Milky Way as suggested 
by the {\it N}-body modelling of BRAVA stellar kinematics; models with 
increased bulge masses are shown to produce larger deviation from 
the data \citep{Shenetal2010}. 

The goal of this paper is to understand the effect of angular momentum of 
a low-mass ICB on the disc instability and kinematics of the final BP bulge. 
In particular, does the cylindrical rotation in BP bulges depend 
on the angular momentum of the ICB and could this possibly explain the deviation from 
cylindrical rotation in observed BP bulges?
Here, we study the formation and growth of a bar and 
BP bulge in galaxies with a low mass rotating ICB using high resolution {\it
N}-body simulations. We find that the kinematics of the final BP
bulge varies with time but the long-term asymptotic behaviour of the resulting
BP bulge does not depend on the initial rotation. We show
how cylindrical rotation in BP bulges changes during the secular evolution
and devise a method to quantify it. 

\begin{figure}
\rotatebox{270}{\includegraphics[height=7.0 cm]{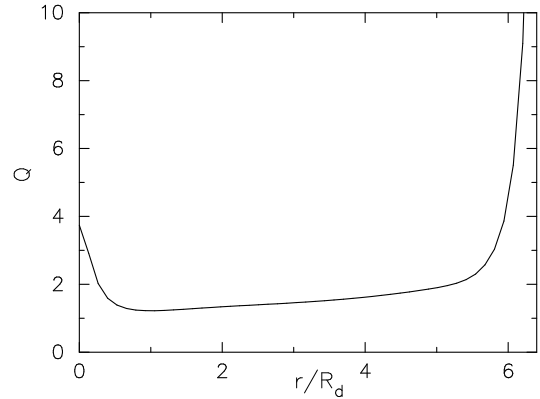}}
\caption{Radial profile of Toomre Q for the initial stellar disc.}
\label{fig:tQ}
\end{figure}
   
The paper is organized as follows. Section~\ref{sec:modelsetup} outlines the 
initial galaxy model and set up for the {\it N}-body simulation. Bar growth and
the formation of BP bulges are described in section~\ref{sec:boxy}.
Kinematics of boxy bulges are described in section~\ref{sec:bkin}. 
Conclusions are presented in section~\ref{sec:discussion}. 

\begin{figure}
\rotatebox{270}{\includegraphics[height=7.5 cm]{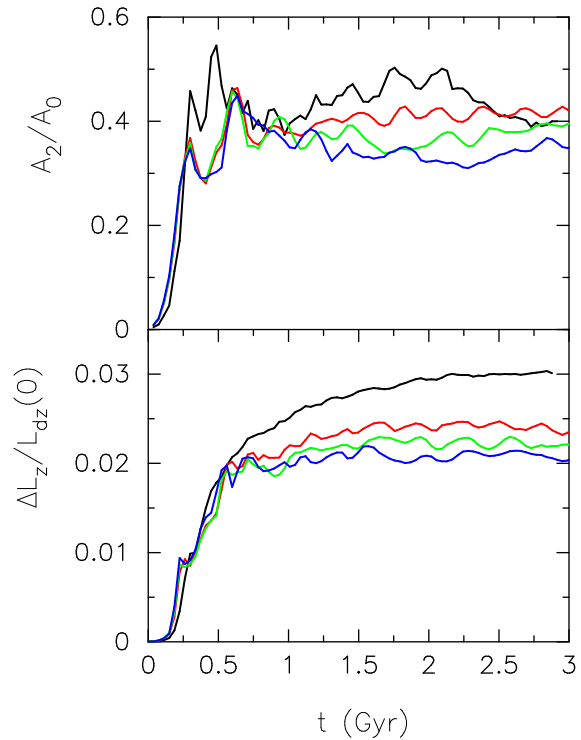}}
\caption{Top panel showing the time evolution of the bar strength
in galaxy models with a preexisting spinning classical bulge.
Bottom panel shows the net gain of angular momentum by the preexisting
classical bulge. $L_{dz}(0)$ denotes the disc specific angular momentum at T=0. 
Black lines in both panels denote the model with non-rotating ICB (RCG004-0). 
Red (RCG004-A), green (RCG004-B) and blue (RCG004-C) lines in both panels indicate 
models with rotating ICBs.}
\label{fig:A2lzb}
\end{figure}

\section{galaxy models with initially rotating classical bulges}
\label{sec:modelsetup}
Equilibrium models of galaxy with ICBs are constructed
using the self-consistent method of \citet{KD1995}. We construct here three galaxy
models with rotating ICBs and one with
non-rotating ICB RCG004-0 taken from \cite{Sahaetal2012} for comparison. Each of 
these initial models consists of a live disc, 
halo and bulge. The initial disc has an exponentially declining surface density with a
scale-length $R_d$, mass $M_d$ and Toomre $Q = 1.4$ at $2.5 R_d$ (see Fig.~\ref{fig:tQ} 
for the radial variation). The live dark
matter halo is modelled with a lowered Evans model and the ICB with a King
model and general details about the model parameters can be found in \citep{Sahaetal2010, Sahaetal2012}. The initial parameters for disc, halo and ICB are identical to model
RCG004 of \cite{Sahaetal2012} and the circular velocity curve is the same as presented in
Fig.~1 of \cite{Sahaetal2012}. The ICBs are strongly flattened by the disc
potential \citep{BarnesWhite1984}. The ICB in the constructed models has
a total mass $M_b = 0.066 M_d$ and the initial ellipticity in edge-on
projection is $\epsilon_b =0.465$.  

\begin{figure*}
\rotatebox{0}{\includegraphics[height=7.0 cm]{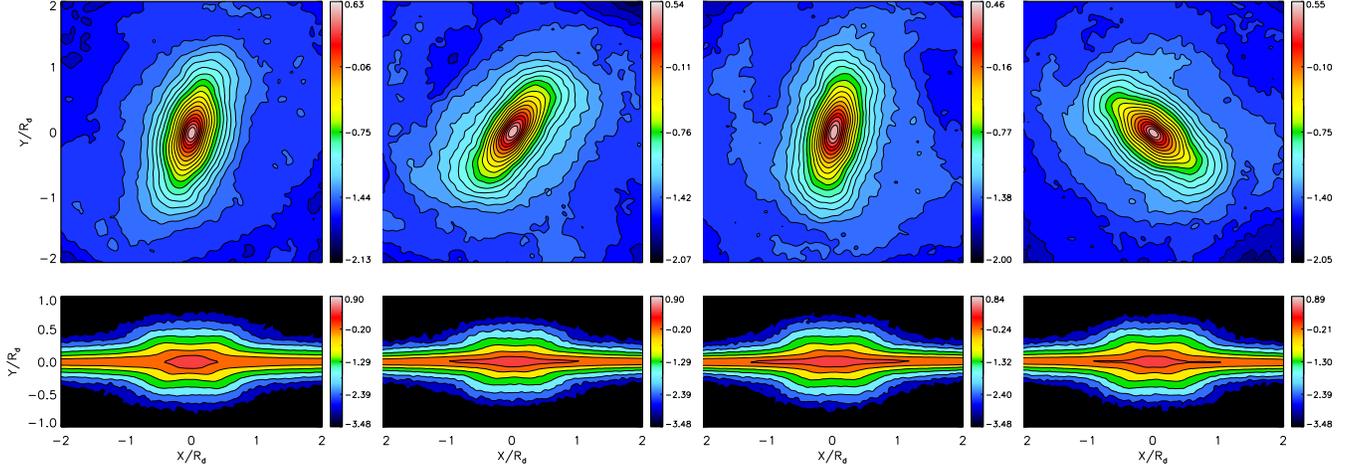}}
\caption{Bars and boxy/peanut bulges at the end of $2.9$~Gyrs. From left to
right, the models are RCG004-0, RCG004-A, RCG004-B and RCG004-C. Initial
rotation of the classical bulge increases from left to right. Bottom panels show
side-on projections of the models, above, along the short axis in the galactic
plane.}
\label{fig:denmap00}
\end{figure*}

The rotating flattened models of ICBs are constructed by 
reversing the velocities of particles with negative angular momenta, which 
remains a valid solution of the collisionless Boltzmann equation 
\citep{Lynden-Bell62}. In this way, three models with rotating ICBs which we
denote as RCG004-A, RCG004-B and RCG004-C are constructed. Their initial velocity
 contours are shown in section~\ref{sec:bkin}. 
Model RCG004-A were constructed by randomly changing the corotating fraction
(i.e., a fraction of bulge orbits corotating with the disc) to $0.75$ (note, corotating
 fraction = 0.5 generates a non-rotating bulge) and it shows some amount of 
cylindrical rotation in the central region, although the ICB is axisymmetric
 \citep{Rowley1988}. 
Models RCG004-B and RCG004-C were constructed slightly
differently based on the work of \cite{DehnenGerhard1993}, in that we reversed the
velocities of bulge stars belonging to a certain range of angular momenta, especially
avoiding the meridional loop orbits, see Fig.~3 of \cite{DehnenGerhard1993}.  

The initial $(V_m/\bar{\sigma})^{*}$ values for all the model bulges are shown 
in Table~\ref{tab:paratab}, where $V_m$ is the maximum 
velocity and $\bar{\sigma}$
is the average velocity dispersion inside the bulge half mass radius. 
The $*$ sign denotes that values are normalized to the corresponding oblate 
isotropic rotator model \citep{Binney1978}.

We scale the models such that $M_d = 4.58 \times 10^{10} M_{\odot}$ and $R_d
=4.0$ kpc. Then the time unit is given by $24.9$ Myr. We use a total of $6
\times 10^{6}$ particles to simulate RCG004-A, RCG004-B and RCG004-C whereas for RCG004-0
 a total of $10 \times 10^{6}$ particles were used. The softening lengths for the disc, 
bulge and halo particles used are $12$, $5$ and $20$ pc respectively 
following the suggestion of \cite{McMillan2007}. These simulations are performed
using the Gadget code \citep{Springeletal2001} with a tolerance parameter
$\theta_{tol} =0.7$,  integration time step $\sim 0.4$ Myr. All the models
are evolved for a time period of $\sim 3.0$ Gyr.

\section{Size and growth of bars and boxy/peanut bulges}
\label{sec:boxy}
Non-linear growth of a bar depends on the amount of angular momentum 
transferred from the inner region of the stellar disc to the outer disc, 
to a classical bulge (if exists)
 and to the surrounding dark matter halo in the host galaxy, all of which act
like a sink. In this section, we focus on the effect of a rotating ICB
on the size and growth of a bar and the final BP bulge.

All $4$ models described in section~\ref{sec:modelsetup} rapidly form a bar
in the disc.
In the upper panel of Fig.~\ref{fig:A2lzb}, we show the evolution of the bar 
amplitude (defined as the maximum of the $m=2$ Fourier coefficient $A_2$
normalized to the axisymmetric component $A_0$) as a function of time for the 
four models. The net amplitude of the bar is lower in models with classical bulges having higher $(V_m/\bar{\sigma})^{*}$. Such a decrement in the bar amplitude correlates well with the 
transfer of angular momenta to the bulge as shown in the bottom panel of 
Fig.~\ref{fig:A2lzb}. It shows the time evolution of the net gain of specific 
angular momentum normalized to the disc's initial angular momentum. The fast
rotating ICB absorbs less angular momentum after $\sim 1$~Gyr as
compared to the moderately rotating or non-rotating ICB. The most
striking difference is between the fast rotating and non-rotating ICBs; after
$\sim 1$~Gyr, the non-rotating ICB gains $\sim 1.5$ times more angular 
momentum than the fast rotating one (see Fig.~\ref{fig:A2lzb}).
 Note that the linear growth rate of the bar is nearly unaffected 
by the initial rotation of the classical bulge.

Fig.~\ref{fig:denmap00} shows the surface density maps of the 
inner $2 R_d$ regions for all the 4 galaxy models at the end of $2.9$~Gyr. 
In the face-on projection, they show strong bars and in the edge-on projection BP bulges.  
We measure the bar size using two methods: one using the drop 
in the ellipticity profile obtained using the IRAF ellipse fitting task on the 
FITS image (see Fig.~\ref{fig:denmap00}) generated from the particle model. In the 
second, we use the radius at which the corresponding position angle of the bar deviates
 by $\sim 5 \deg$ \citep{Erwin2005}. We quote the bar sizes (denoted by $R_{bar}$) as the minimum of these 
two measurements obtained at the end of $2.9$~Gyrs (see Table~\ref{tab:paratab}). 
For a list of methods on bar size measurement, see \cite{Athanamisi2002}. We have
also used an independent method to derive the bar size directly from the
particle model. In that, we use the FWHM of the radial
profile of the bar amplitude as one measurement and the radius at which the phase angle of
the bar deviates by $5 \deg$ as the second. The minimum of these measurements
is nearly the same as that obtained from the ellipse fitting method. From these
measurements, it is found that the bar sizes are, in general, smaller if the
ICB is rotating. On average, for the low mass ICBs considered here rotating with
 a $(V_m/\sigma)^{*} \sim 0.42 - 0.72$, we found about $10 - 40\%$ reduction in the 
bar size compared to the non-rotating case.

\begin{figure}
\rotatebox{-90}{\includegraphics[height=8.5 cm]{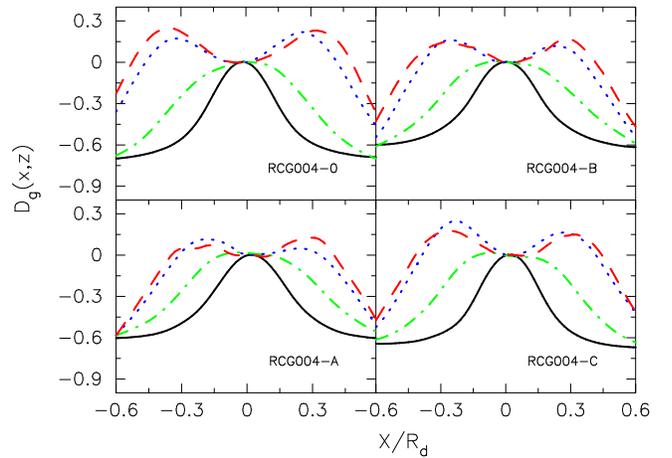}}
\caption{Shape of the function $D_g$ along slits parallel to the major axis
for all the four models.
In each panel, profiles are drawn at $z/R_d$ =0.009 (solid black line), 
0.11 (green dashed-dot line), 0.23 (blue dotted line), and 
0.3 (red dashed line). The profiles at $z/R_d =0.3$ are used here to measure
the size of the boxy/peanut bulges.}
\label{fig:sigLOS}
\end{figure}

The size of a BP bulge is calculated by finding zeros of the 
function $D_g (x,z)$ defined as
 
\begin{equation}
D_g (x,z) = \frac{\Sigma_{los}(x,z) -\Sigma_{los}(0,z)}{\Sigma_{los}(0,z)},
\end{equation}

\noindent for a set of smoothed surface density ($\Sigma_{los}$) profiles (see
Fig.~\ref{fig:sigLOS}) along slits parallel to the major axis ($x$) of a 
galaxy model such as shown in the lower panels of Fig.~\ref{fig:denmap00}.

The shape of the function $D_g$ varies along the vertical direction. Close to
the midplane (i.e., $z =0$) of the galaxy, $D_g$ is convex with the zeros
occurring at $x=0$ (i.e., along the minor axis) and values of $D_g$ are always
negative at $x \ne 0$. Slightly above the midplane (e.g., $z=0.11 R_d$ in
Fig.~\ref{fig:sigLOS}) $D_g = 0$ in the inner regions ($x \sim 0.1 R_d$) of some
models and smoothly goes to negative beyond that. We define such a profile as
characteristic of a boxy shaped bulge. At higher values of
$z$ (e.g., at $z=0.23, 0.3 R_d$, in Fig.~\ref{fig:sigLOS}) the shape of the function
 $D_g$ takes the form of a characteristic peanut shape. Above $z=0.3 R_d$, we
find that shape of the function $D_g$ becomes irregular for some models and
makes it harder to compare them with other models. We consider the
profiles at $z=0.3 R_d$ to compute the length of the BP bulge for all
the models. The length of the BP
bulge (denoted by $BPL$, see \cite{Luttickeetal2000}) is obtained by
calculating the distance between two 
zeros of the function $D_g$ on either side of the galaxy centre and they are 
enlisted in Table~\ref{tab:paratab}. This method reliably returns the 
values of $BPL$ as a function of vertical height ($z$) when there is a clear 
peanut signature and it is found that 
the largest BP bulge forms in model RCG004-0 with non-rotating ICB.
 
The strength of the BP bulge in our models is obtained by calculating
the maximum of the $m=2$ Fourier coefficient (denoted by $C_{2,z}^{max}$) of the
z-coordinates of the disc particles belonging to the bar
\citep{Martinez-Valpuesta2008}. In order to understand the effect of 
ICB on the disc, we have used only disc particles to compute $A_2$ and 
$C_{2,z}$. The values of $C_{2,z}^{max}$ are shown in
Table~\ref{tab:paratab}. 
It is found that the strength of
BP bulges in our models with non-rotating ICB and fast rotating
ICB are nearly the same. However, with moderately or slowly rotating ICBs,
the strength is lowered by $\sim 20 - 40\%$.  

Further, we perform a 2D bulge-disc decomposition on the FITS images produced
from our simulations using GALFIT \citep{Pengetal2002} and  derive the bulge 
half mass radii ($R_{b,{1/2}}$), sersic indices ($n_b$) (see
Table~\ref{tab:paratab}) and ellipticities for all the 
four final BP bulges shown in Fig.~\ref{fig:denmap00}. Note that the BP bulge here is a composite bulge which is a superposition of the BP bulge formed out of the disk material and the ICB at $2.9~Gyrs$. The initial value of $R_{b,{1/2}}$ for the non-rotating ICB is $0.26 R_d$. We do not find any
definite trend either in the sersic indices or bulge sizes with the
initial rotation in the classical bulges but the slowly rotating one in our
model (RCG004-A) forms shorter bulge and have larger sersic index.

\begin{table}
\caption[ ]{Rotating ICBs and properties of bars and boxy/peanut bulges at $2.9$~Gyr in our
simulations. Symbols are explained in section~2 and 3.}
\begin{flushleft}
\small\addtolength{\tabcolsep}{-2pt}
\begin{tabular*}{8.5cm}{cccccccccc}  \hline\hline 
Galaxy   &$(V_m/\sigma)^{*}$ & $R_{bar}$ & $R_{{b},{1/2}}$ & $n_b$ &
$BPL$ & $C_{z,2}^{max}$ &\\
models & (at T=0)& ($\times R_d$)    & ($\times R_d$) & & ($\times R_d$)  &  &  \\
\hline
\hline
\vspace{0.1cm}

\vspace{0.1cm}RCG004-0   & 0.0  &1.35 & 0.32 &1.12 & 1.02 & 0.085 \\

\vspace{0.1cm}RCG004-A   & 0.42 &0.95 & 0.25 &1.33 & 0.74 & 0.056 \\

\vspace{0.1cm}RCG004-B   & 0.52 &1.20 & 0.27 &1.13 & 0.83 & 0.068\\
RCG004-C   & 0.72 &1.10 & 0.33 &1.03 & 0.86 & 0.084\\

\hline
\end{tabular*}
\end{flushleft}
\label{tab:paratab}
\end{table}

\section{Kinematics of boxy/peanut bulges}
\label{sec:bkin}
Kinematically BP bulges are distinguished from the classical bulges in 
that BP bulges possess cylindrical rotation as observed in a number of external 
galaxies \citep{KI1982,Falcon-Barrosoetal2006} as well as in our Milky Way \citep{Howardetal2009} but see
\cite{Williamsetal2011} for exceptional cases which displayed noticeable deviation from cylindrical rotation. All four
models of BP bulges, studied here, rotate cylindrically but the degree of
cylindrical rotation is seen
to be varying with time as the model galaxy evolves. In section~\ref{sec:cylin},
we provide a simple prescription to quantify the degree of cylindrical
rotation. The current section is focused on understanding how the kinematics of 
the BP bulges depend on the initial rotation in the ICBs. 

\begin{figure}
\rotatebox{-90}{\includegraphics[height=8.0 cm]{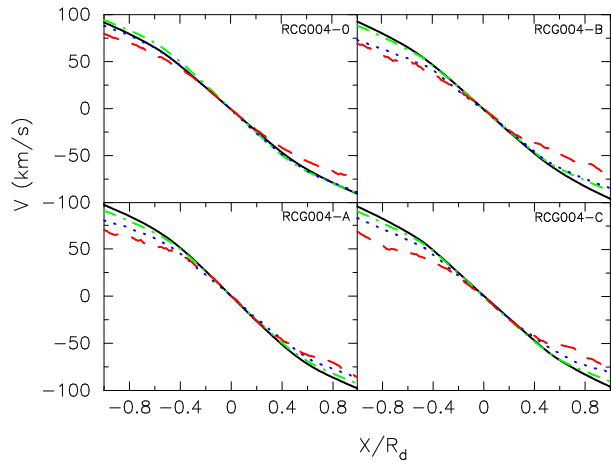}}
\caption{ The mean LOS velocity profiles along 4 different slits placed at 
$z/R_d$ =0.009 (solid black line), 0.11 (green dashed-dot line), 0.23 (blue
dotted line), and 0.35 (red dashed line) are shown in each panel. All the
velocity profiles are shown for $T=2.9$~Gyr. Cylindrical rotation is evident within
the bulge half-mass radii of all models and extends beyond
$R_{b,{1/2}}$ for the model with non-rotating ICB (RCG004-0).}
\label{fig:vmajor0}
\end{figure}

\begin{figure}
\rotatebox{0}{\includegraphics[height=10.0 cm]{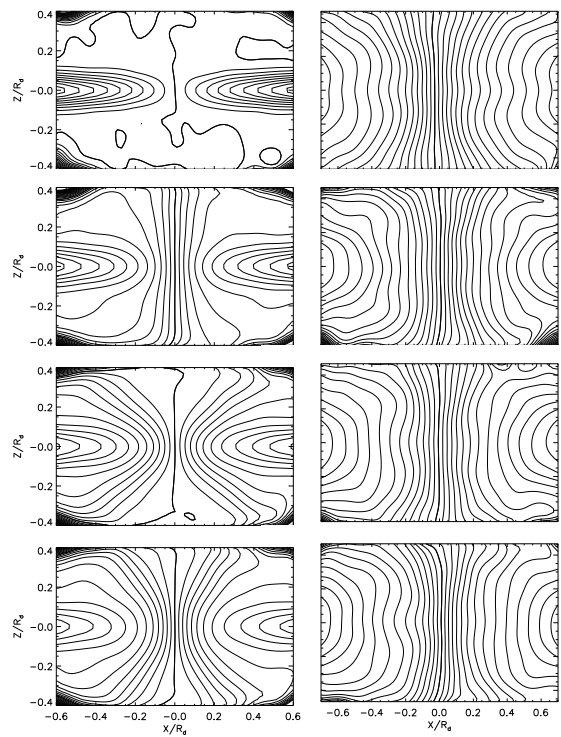}}
\caption{Spider diagrams for the mean LOS velocity of stars in the
boxy bulge region which includes stars from the disk as well. From top to bottom, panels are drawn for models RCG004-0, RCG004-A,
RCG004-B and RCG004-C. From left to right panels, they show 
evolution of the velocity structure between initial (T=$0$) and final states 
(T=$2.9$~Gyr).}
\label{fig:cont77}
\end{figure}

In Fig.~\ref{fig:vmajor0}, we show mean LOS velocity profiles along slits placed parallel
to the major axis of our galaxy model to probe how the rotation of a BP
bulge varies along the vertical direction. At the end of $2.9$~Gyrs, all the
models exhibit clear (see section~\ref{sec:cylin} for numbers) cylindrical
rotation up to a height of $\sim 0.23 R_d$ ($\simeq R_{{b},{1/2}}$) which is about 
$900$ pc for the adopted scaling above the galactic midplane. 
Unlike models with rotating ICBs, the model with non-rotating ICB (RCG004-0) show
cylindrical rotation up to a height of $\sim 1200$ pc above the midplane. This is 
also clear from the 2D velocity
contours of the BP bulge region as depicted in Fig.~\ref{fig:cont77}. In
models with rotating ICBs, the velocity contours start deviating above a
height of $\sim 0.25 R_d$. Fig.~\ref{fig:cont77} reveals that, although the initial bulge rotation for all the
four models are quite different, at the end of $2.9$~Gyrs they show similar
kinematics. It remains to verify whether their evolutionary paths
are also nearly identical. We address this issue in section~\ref{sec:cylin}.   

\begin{figure}
\rotatebox{-90}{\includegraphics[height=7.5 cm]{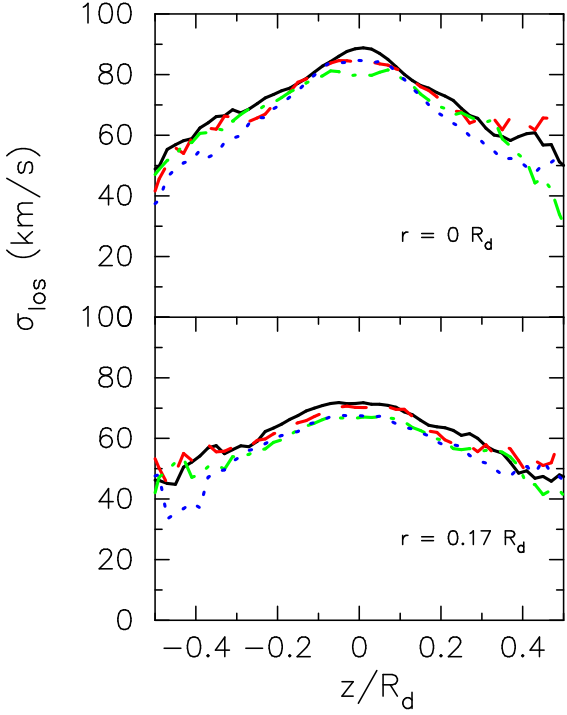}}
\caption{The vertical profiles of LOS velocity dispersion at
different locations from the centre of the galaxy. The upper panel shows the
dispersion profiles along the minor axis while the bottom panel shows for $X=0.17 R_d$.
In both panels, solid black lines denote RCG004-0, green dash-dot lines denote
RCG004-A, blue dotted lines RCG004-B and red dashed lines RCG004-C. All the
dispersion profiles are derived at $T=2.9$~Gyr.}
\label{fig:veldisp}
\end{figure}

In Fig.~\ref{fig:veldisp}, we show the variation of mean velocity dispersion ($\sigma_{los}$)
 of stars in the final BP bulge along the vertical direction at two different radii
from the centre of the respective galaxy model. 
In all the models, the minor axis $\sigma_{los}$ profiles show steeper vertical gradient
than the profiles at $X = 0.17 R_d$ (which is well within the bulge
half-mass radii). A vertical gradient in the stellar velocity
dispersion is observed in classical bulges as well as in boxy bulges
\citep{Williamsetal2011}. It has also been observed that the velocity dispersion
is remarkably constant along vertical direction in
NGC4594 (a classical bulge) and in the well known boxy bulge of NGC4565
\citep{KI1982}. So a gradient in the velocity dispersion does not necessarily
indicate the nature of the underlying bulge. If measured in the boxy bulge
region, one would perhaps expect a shallow gradient in the velocity dispersion.
From Fig.~\ref{fig:veldisp}, it is clear that the rotation of the low mass ICBs
 does make little difference in the velocity dispersion of the final BP bulge.  

\subsection{Criteria for cylindrical rotation}
\label{sec:cylin}
By definition, cylindrical rotation means that the mean LOS velocity
$V_{los}(x,z)$ of stars at a particular distance ($x$) from the centre is 
independent of its vertical height ($z$) i.e., zero vertical gradient 
($dV_{los}(x,z)/dz = 0$). This is 
approximately true for observed barred galaxies as well as for bulges in N-body simulations
\citep{Combesetal1990, Athanamisi2002}. Here, we construct 
a simple criterion for measuring the degree of cylindrical rotation in BP bulges.

We consider again a set of slits, nearly covering the vertical extent of the BP bulge, parallel to the major axis in edge-on
projection. Along these slits, we have the LOS surface density
$\Sigma_{los}(x,z)$ and velocity $V_{los}(x,z)$ of stars.
For a given slit placed at $z = z_i$, we can compute the mass weighted 
line-of-sight velocity for all the stars within a projected distance $X_j$
from the minor axis of the edge-on image. This quantity may be normalized by
replacing the individual velocities of all the stars within $X_j$ with a
velocity at the bulge half-mass radius close to the disc midplane
$V_{los}(R_{{b},{1/2}},z\simeq 0)$. After some experimenting, we adopt a
weighting scheme and define a dimensionless quantity for the given slit as
follows: 

\begin{equation}
\delta_{CL}(z,X_j) = \frac{\int_0^{X_j}{\Sigma_{los}(x,z) V_{los}(x,z_i) x^2
dx}}{V_{los}(R_{{b},{1/2}},z\simeq 0) {\int_0^{X_j}{\Sigma_{los}(x,z) x^2 dx}}}.
\label{eq:delCL}
\end{equation}

\begin{figure}
\rotatebox{-90}{\includegraphics[height=8.5 cm]{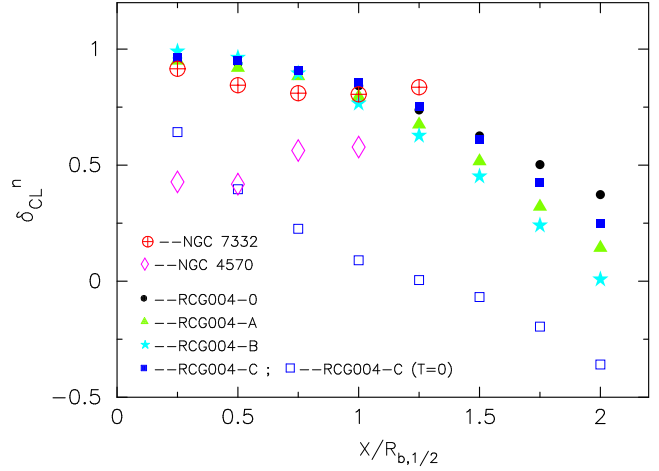}}
\caption{Degree of cylindrical rotation at different major axis 
radii of our N-body models of BP
bulges plotted the end of 2.9 Gyrs. Overplotted here are the degree of
cylindrical rotation for two galaxies NGC7332 and NGC4570.}
\label{fig:xdeltaCL}
\end{figure}

\begin{figure}
\rotatebox{-90}{\includegraphics[height=8.5 cm]{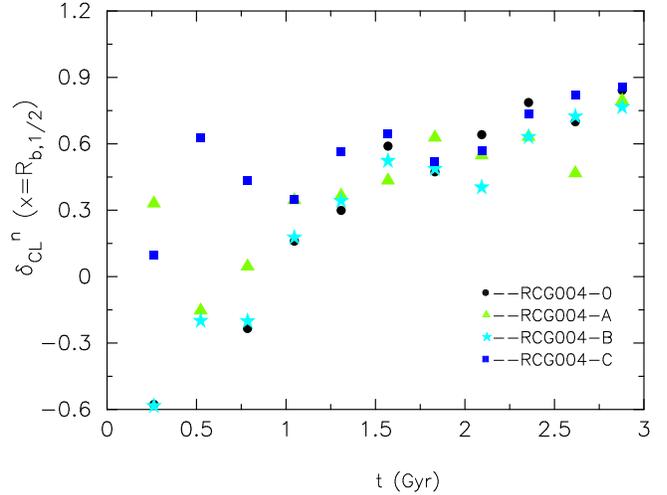}}
\caption{Time evolution of the degree of cylindrical rotation calculated at the
bulge half mass radii for four N-body models.}
\label{fig:tdeltaCL}
\end{figure}

\noindent We compute this quantity for the set of slits considered. 
The adopted weighting scheme puts
more weight on the data points further from the minor axis. Note that, both the
numerator and denominator  have the dimension of angular momentum. If the stars
close to the disc midplane rotate faster than the stars above, the denominator
will always be greater than the numerator. If we had the knowledge of
deprojection, one could interpret the above
formulae as how the average rotational motion of stars associated with an 
infinitesimally thin disc at a vertical distance $z=z_i$ compares to 
the one close to the midplane. In general, the values of $\delta_{CL}$ are 
$\sim 1$ near the midplane and decrease as $z$ increases. 

To proceed further, we plot the computed $\delta_{CL}$ values against the slit
location ($z$) above the galactic midplane.
The velocities of stars decrease along the vertical direction in our model
galaxies and the same is expected for galaxies in equilibrium. So we expect 
$\delta_{CL}$ values can
be well approximated by a linear function of $z$, where $z$ is scaled to $R_{{b},{1/2}}$. We then fit a straight line to
the ({$\delta_{CL}$ , $z$}) curve and derive the slope $m_{CL}$. The degree of
cylindrical rotation in the BP bulge is then defined as

\begin{equation}
\delta_{CL}^n (X_j) = 1 + m_{CL}.
\end{equation}  

\noindent The values of $\delta_{CL}^n$ are generally less than $1$ unless
$\delta_{CL}$ increases in the vertical direction which is unlikely for
realistic models of galaxies. Note $\delta_{CL}^n$ is undefined for a
non-rotating bulge because Eq.~\ref{eq:delCL} would give $\frac{0}{0}$.
$\delta_{CL}^n = 1$ denotes pure cylindrical rotation. The degree of cylindrical
rotation decreases as the value of $\delta_{CL}^n$ decreases from unity. One
could carry out the above exercise for different values of $X_j$ (e.g., $X_j =
0.5 R_{{b},{1/2}}, R_{{b},{1/2}}, 2 R_{{b},{1/2}}$ etc.) and show how the degree of cylindrical rotation in a BP bulge would depend on the chosen value of $X_j$. 

In Fig.~\ref{fig:xdeltaCL}, we show how the degree of cylindrical rotation
varies within the BP bulge at the end of $T = 2.9$~Gyrs. Within $R_{{b},{1/2}}$,
all the model BP bulges show nearly the same degree of cylindrical rotation.
Beyond the bulge effective radii, $\delta_{CL}^n$ decreases rapidly. For the
models with moderately rotating bulges (e.g., RCG004-A and RCG004-B),
$\delta_{CL}^n$ values are systematically lower (see Fig.~\ref{fig:xdeltaCL}). 
We have also plotted the values of $\delta_{CL}^n$  for the model
RCG004-C at $T= 0 $ when the bulge showed no sign of cylindrical rotation (see
Fig~\ref{fig:cont77}). 

In order to test the above formulae, we applied it to two real galaxies, NGC7332
and NGC4570. NGC7332 is a well studied galaxy in the literature and shows
prominent cylindrical rotation \citep{Falconetal2004}. We have used SAURON
integral-field kinematics data \citep{Emsellemetal2004} for the 
above two galaxies. The bulge effective radii ($R_{b,{1/2}}$) 
for NGC4570 and NGC7332 were taken to be $14$ and $6$ arcsec respectively 
\citep{Emsellemetal2004, Fisheretal1994}. Our analysis indicates that NGC7332 
has a high degree of cylindrical rotation($\delta_{CL}^n \ge 0.8$) within the bulge 
effective radius. On the other hand, NGC4570 shows non-cylindrical rotation within 
its bulge effective radius (see Fig.~\ref{fig:xdeltaCL}). 

Based on the above measurements of the degree of cylindrical rotation 
in about $6$ 
bulges, we introduce $\delta_{crit}=0.75$ as the boundary between cylindrical 
and non-cylindrically rotating bulges. Note that such a demarcation is not based 
on any rigorous calculation but should be treated as an operational definition. $\delta_{crit}=0.75$ separates well, 
the rotating classical bulge (RCG004-C (T=0)) and non-cylindrically rotating 
spheroidal bulge of NGC 4570 from the rest of our sample. So a model bulge with 
$\delta_{CL}^n > \delta_{crit}$ ($\delta_{CL}^n < \delta_{crit}$) would be 
described as a {\it cylindrical rotator} (otherwise). It would be interesting to 
investigate $\delta_{CL}^n (X=R_{{b},{1/2}})$ for a larger sample of boxy bulges.

\subsection{Time dependent cylindrical rotation}
In Fig.~\ref{fig:cont77}, we showed that the kinematics of BP bulges at the end
of $2.9$~Gyrs turned out to be similar and independent of the initial rotation
of the ICBs. Here, we compute the degree of cylindrical rotation within the
bulge effective radius as a function of time for all the four models and plotted
them in Fig.~\ref{fig:tdeltaCL}. This figure reveals interesting evolutionary behaviour for
the BP bulges. First of all, Fig.~\ref{fig:tdeltaCL} shows clearly that the
degree of cylindrical rotation varies with time and that there are substantial
differences at early times depending on the initial rotation of the ICBs.
Nevertheless, irrespective of the initial condition, each model BP bulge evolves
towards attaining a higher degree of cylindrical rotation and within their
respective bulge effective radii, the values of $\delta_{CL}^n$ approach the
same level. From Fig.~\ref{fig:tdeltaCL}, we notice that it takes 
about $2.0$~Gyr
since the buckling episode for the cylindrical rotation to be present at the level
of $\delta_{CL} \ge 0.75$ in our model bulges. 
In the section below, we show a snapshot during the evolution 
at $T = 1.05$~Gyr when cylindrical rotation in the bulge is still developing.

\subsection{Boxy/peanut bulges with non-cylindrical rotation}
Non-cylindrically rotating boxy bulges are seen in some galaxies e.g., NGC1381,
NGC5746 and IC4767 \citep{Williamsetal2011}. It is not clear what might have
caused such boxy bulges to rotate non-cylindrically. 

\begin{figure}
\rotatebox{0}{\includegraphics[height=8.0 cm]{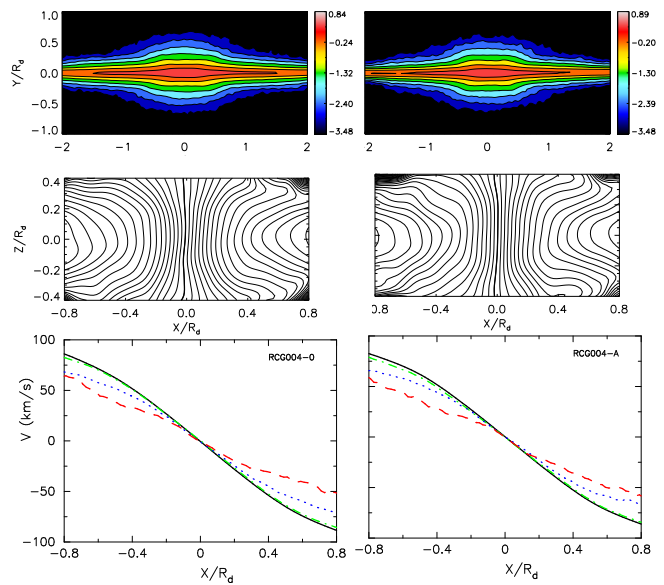}}
\caption{Top panels showing the surface densities for two models RCG004-0 (left)
and RCG004-A (right) at T=1.05 Gyr. The middle panels show the contours 
of the 2D mean LOS velocity field for both the models. Velocity profiles at 4
different slits placed at $z/R_d = 0.0$(black solid line), $0.11$ (green dash-dot
line), $0.23$ (blue dotted line) and $0.35$ (red dashed line) parallel to the
major axis of the galaxy are depicted in the bottom panels. Both models show
deviations from pure cylindrical rotation within their respective bulge half-mass
radii.}
\label{fig:dcontvel0A}
\end{figure}

\begin{figure}
\rotatebox{0}{\includegraphics[height=8.0 cm]{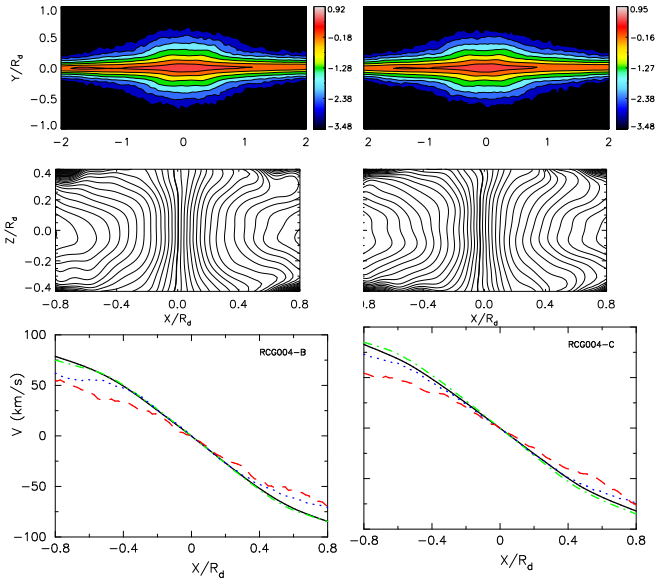}}
\caption{ Same as in Fig.~\ref{fig:dcontvel0A} but for models RCG004-B (left)
and RCG004-C (right). }
\label{fig:dcontvelDC}
\end{figure}

The analysis of the previous section demonstrate that the kinematics of
BP bulges are time-dependent, especially the
degree of cylindrical rotation. From Fig.~\ref{fig:tdeltaCL}, we see that at $T
= 1.05$~Gyr beyond the buckling instability phase, two of the model bulges
have less cylindrical rotation than the other two. In Fig.~\ref{fig:dcontvel0A}
and Fig.~\ref{fig:dcontvelDC}, we show the surface density and kinematics of
all four model galaxies at $T= 1.05$~Gyr. According to our convention for 
$\delta_{crit}$, all $4$ models of BP bulges at this time are non-cylindrically rotating. 
Note however, that the BP bulge in RCG004-C (see Fig.~\ref{fig:dcontvelDC}) has 
cylindrical rotation within a smaller region about the minor axis. Since 
$\delta_{CL}^n(X=R_{{b},{1/2}})$ refers to the bulk of the bulge, such small 
details are averaged out. The present analysis 
indicates that boxy bulges with non-cylindrical rotation could still be in their 
early phase of evolution i.e., have formed relatively recently in their host 
galaxies. Another possibility is that in those galaxies the ICB was too massive 
to be spun-up all the way to cylindrical rotation; the response of massive ICBs 
to a growing bar will be investigated in a future publication. 

As shown in section~\ref{sec:boxy}, all the rotating and non-rotating ICBs gain 
angular momentum emitted by the bar. As a result of the angular momentum gain, 
the ICBs spin up and their orbital structures evolve over time. A spherical
harmonics analysis of the 3D density distribution of the ICBs showed that their
inner regions formed a bar-like structure. For details on the case of 
non-rotating ICB (here, model RCG004-0), the readers are referred to \cite{Sahaetal2012}.
The inner regions of these ICBs exhibit cylindrical rotation beyond about
$0.5$~Gyr which marks the bar buckling instability phase in the disc.
The net cylindrical rotation in the composite bulges (e.g., shown in
Fig.~\ref{fig:denmap00}, Fig.~\ref{fig:dcontvel0A} or Fig.~\ref{fig:dcontvelDC}) 
has its contribution from the evolved rotating ICB and the
boxy bulge that form out of the bar buckling instability.

\section{Conclusions}
\label{sec:discussion}
We have investigated the morphology and kinematics of
a BP bulge that results from the instability of an axisymmetric disc assembled around 
a rotating low-mass ICB during the galaxy formation process. The composite 
BP bulge in our
model contains a rotating ICB and shows highly evolving kinematics. Our main conclusions 
are highlighted below:

1. The size and strength of the bars and resulting BP bulges in our simulations are
reduced by $\sim 10 - 40 \%$ in galaxy models containing ICBs rotating with
$(V_m/\sigma)^{*} = 0.4 - 0.7 $ compared to one with $(V_m/\sigma)^{*} = 0.0$ .

2. Rotating ICBs gain less angular momentum compared to the non-rotating ICB
during the secular evolution.

3. We show that cylindrical rotation in BP bulges in our simulations are
time-dependent. During the
early phases of BP bulge formation, significant deviations from cylindrical
rotation can be observed. We quantify such deviations by providing a simple formulae. 
Our formulae when applied to NGC7332, gives a value for
$\delta_{CL}^n =0.8$ within the bulge half-mass radius. Using the same formulae,
we confirm that NGC4570 is not cylindrically rotating.

4. The degree of cylindrical rotation in the composite BP bulges does not depend on
the angular momentum of the ICBs sufficiently long time after its
formation.
 
5. The early kinematics of BP bulges with rotating low-mass ICBs in our simulations suggest that the BP bulges with deviation from cylindrical rotation might have formed relatively recently.
 
\section*{Acknowledgement} 
\noindent KS acknowledges support from the Alexander von Humboldt Foundation and visiting
 support from IUCAA, Pune. 
The authors thank the referee Dr. Juntai Shen for constructive comments on the manuscript.
The authors would like to thank Michael J. Williams, Chaitra Narayan for useful
comments and discussion.


\end{document}